\documentclass[11pt]{article}
\usepackage{amssymb,amsmath}
\usepackage{graphicx}
\usepackage{color}
\usepackage{authblk}
\begin{document}

\title{The lifetime of cosmic rays in the Milky Way}

\author{Paolo Lipari}

\affil{\footnotesize 
INFN, sezione di Roma, Piazzale Aldo Moro 2, 00185 Roma, Italy.
\\
Gran Sasso Science Institute (INFN), viale F. Crispi 7, 67100 L'Aquila, Italy
}

\date{19 july 2014}

\maketitle

\begin{abstract}
The most reliable method to estimate the residence time of 
cosmic rays in the Galaxy is based on the study of 
the suppression, due to decay,
of the flux of unstable nuclei such as beryllium--10,
that have lifetime of appropriate duration. 
The Cosmic Ray Isotope Spectrometer (CRIS) collaboration
has measured the ratio between the fluxes
of beryllium--10 and beryllium--9 
in the energy range $E_0 \simeq 70$--145~MeV/nucleon, and 
has used the data to estimate an escape time 
$\tau_{\rm esc} = 15.0 \pm 1.6$~Myr.
This widely quoted result 
has been obtained in the framework of 
a simple leaky--box model 
where the distributions of escape time and age for stable particles 
in the Galaxy are identical and have exponential form.
In general, the escape time and age distributions 
do not coincide,  they are not unique (because they depend
on the injection or observation point), and 
do not have a simple exponential shape. It is  therefore necessary 
to discuss the measurement of the beryllium ratio
in a framework that is more general and more
realistic than the leaky--box model.

In this work we compute the escape time and age distributions
of cosmic rays in the Galaxy in a model
based on diffusion that is much more realistic than the simple leaky--box,
but that remains sufficiently simple to have exact analytic solutions.
Using the age distributions of the model to interpret the measurements of
the beryllium--10 suppression, one obtains a 
cosmic ray residence time that is significantly longer 
(a factor 2 to 4 depending on the extension 
of the cosmic ray halo) than the leaky--box estimate.
This revised residence time implies a proportional reduction of the power
needed to generate the galactic cosmic rays.
\end{abstract}

\section{Introduction}
\label{sec:introduction}
The average residence time of cosmic rays (CR) in 
the Milky Way is a very important quantity in high energy astrophysics,
and is a key element to determine the power required to generate 
the galactic CR.
The most direct way to estimate the average residence time
is the measurement of the suppression, due to radioactive decay, 
of the flux of an unstable nucleus that has a lifetime
comparable with the residence time. A longer residence time
obviously implies a larger decay probability and a smaller flux.
The comparison of the fluxes of two isotopes of the same
chemical element, one stable and the other unstable 
allows to measure the decay suppression for the unstable
particle, and the result can then be used to estimate the lifetime 
of the CR particles.

The most attractive element to perform this program is beryllium
that has two stable isotopes (${}^{7}$Be and ${}^{9}$Be)
and one unstable (${}^{10}$Be)
with half--life $1.51\pm 0.04$~Myr \cite{Tilley:2004zz}.
Beryllium is a very rare element in ordinary matter,
and essentially all beryllium nuclei in the cosmic rays
have not been directly accelerated, but are ``secondaries''
formed by the fragmentation of heavier nuclei, mostly carbon and oxygen,
as they interact with the interstellar gas.
This implies that the injection rates for the different isotopes
can be calculated from a knowledge of the fluxes of the primary nuclei 
and of the relevant fragmentation cross sections.
More explicitely one can write the injection rate (at the energy per nucleon
$E_0$ and the space point $\vec{x}$) of the isotope $j$ in the form:
\begin{equation}
q_j (E_0, \vec{x}) = n_{\rm ism} (\vec{x}) \;\beta \, c \; \sum_A n_A (E_0, \vec{x}) 
~\sigma_A (E_0) \, B_{j\to A} (E_0) ~,
\label{eq:injection}
\end{equation}
where $n_{\rm ism} (\vec{x})$ is the density of the interstellar medium gas,
$n_A(E_0, \vec{x})$ is the density of CR nuclei of 
type $A$ at the same energy per nucleon, $\sigma_A$ is the
total charge changing cross section for nucleus $A$,
and $B_{A\to j}$ the fraction of interactions 
where a nucleus of type $j$ is produced.
The summation is in principle extended to all nuclei 
of sufficiently large mass, but in practice it is dominated
by the contributions of carbon and oxygen.
In equation (\ref{eq:injection}) we have also used the (good) 
approximation that the nuclear fragments emerge from the
interaction with the same velocity 
as the primary nucleus, or equivalently with the same energy per nucleon.

The ratio between the fluxes of beryllium--10 and beryllium--9 
at a fixed energy per nucleon (left implicit in the notation)
can then be written as the product of three factors: 
\begin{equation}
r = \frac{\phi_{10}}{\phi_{9}} =
\left \langle P_{\rm surv} \right \rangle \; r_\sigma \; r_{\rm prop} = 
\left \langle P_{\rm surv} \right \rangle
~ 
\left (\frac{\sum_A \phi_A \; \sigma_A \; B_{A \to 10}}
{\sum_A \phi_A \; \sigma_A \; B_{A \to 9}} \right ) \; r_{\rm prop}~.
\label{eq:ber-ratio}
\end{equation}
In this equation the factor $r_\sigma$ 
is the ratio of the injections for the two isotopes calculated 
using equation (\ref{eq:injection}) and the assumption
that the shape of the energy spectra for the different nuclear types
is independent from the space coordinates. The factor $r_{\rm prop}$ takes
into account the fact that, even neglecting the effects of decay
for beryllium--10, the propagation properties 
of the two isotopes are not identical.
This is because nuclei of the two isotopes with the same energy 
per nucleon have rigidities that differ by a factor 10/9,
and their absorption due to interaction with the interstellar gas 
are not identical because of small differences in their cross sections.
Finally, the factor $\langle P_{\rm surv} \rangle$ takes into account the 
effect of decay for the flux of the unstable nucleus.

Assuming that the energy of a cosmic ray nucleus
remains constant after the injection, and using the notation 
$f_{\rm age}(t)$ for the (normalized) distribution of 
the time $t$ elapsed between 
the instants of injection and observation of a particle,
the average survival probability can be calculated as:
\begin{equation}
\left \langle P_{\rm surv} \right \rangle = 
\int_0^\infty dt ~ f_{\rm age} (t) ~e^{-t/T_{\rm dec}}
\label{eq:def0}
\end{equation}
where $T_{\rm dec}= \tau_{\rm dec} \, \gamma$ is the decay time of
the isotope (with $\gamma$ the Lorentz factor of the particles
at the energy under study).
The distribution $f_{\rm age} (t)$ must be 
calculated {\em neglecting decay}.

Measurements of the beryllium ratio for nuclei
with kinetic energy in the interval 70--150~MeV per nucleon have
been obtained by different experiments
\cite{beryllium1,beryllium2,beryllium3}, and the results have been used
to estimate the CR confinement time. 
The Cosmic Ray Isotope Spectrometer (CRIS) collaboration aboard the
Advanced Composition Explorer spacecraft \cite{beryllium3} 
has measured the beryllium ratio 
$r= 0.123 \pm 0.013$,
$0.115 \pm 0.013$ and
$0.122 \pm 0.016$ for nuclei in the kinetic energy intervals
[70--95], 
[95-120] and
[120--145]~MeV/nucleon. The CRIS collaboration has estimated the product
$r_\sigma \, r_{\rm prop}$ as approximately unity and, 
interpreting the result on the basis of a steady state leaky--box model,
has obtained an ``escape time'' $\tau_{\rm esc} = 15.0 \pm 1.6$~Myr.
In the leaky--box model
the cosmic rays age and escape time distributions 
are identical and have an exponential 
form that is determined by the single parameter 
$\tau_{\rm esc}$ that has the physical meaning of the (position independent) 
average age and average escape time of the particles.

The cosmic rays  ``age'' and  ``escape time''  are distinct concepts,
the first one measures the time elapsed between the instants of injection
and observation of a particle, while the second one  is the total
time that a particle spends in the Galaxy after injection.
Accordingly, the age and escape time distributions are 
in general not identical,
and they are also not unique: the age distribution
depends on the point of observation, and 
the escape time distribution depends on the point of
injection.
It is intuitive that particles injected
near the center (periphery) of the Galaxy need a longer (shorter) time
to escape, and the average age of the particles observed near (far) from
the CR sources is shorter (longer).
An interpretation of the observations of beryllium--10 and other unstable
nuclei that goes beyond the leaky--box approximation
should address these ambiguities.
The suppression of the flux of an unstable particle,
as shown in equation (\ref{eq:def0}), depends on the age distribution
at the observation point.

A fundamental point is that to interpret a measurement of 
$\langle P_{\rm surv} \rangle$ to obtain information
about the cosmic ray residence time 
(for example to estimate $\langle t_{\rm age} \rangle$ at the 
solar system position) it is necessary
to make some assumption about the shape of the age distribution.
The leaky--box hypothesis that the age distribution
has a simple exponential form has not a good physical motivation, 
and uncertainties about the shape of the age distribution, 
could very well be (see discussion below) 
the dominant source of error in the estimate of the CR galactic residence time. 
It is clearly necessary to consider 
models of the age distributions that are 
more realistic and have a more robust 
physical motivation.

In the standard analysis, based on the leaky--box model
and used in \cite{beryllium1,beryllium2,beryllium3},
it is assumed that the age time distribution is a simple exponential:
\begin{equation}
f_{\rm age} (t) = \frac{1}{\langle t_{\rm age}\rangle} e^{-t/\langle t_{\rm age} \rangle} ~.
\label{eq:l-box}
\end{equation}
The integral in equation (\ref{eq:def0}) can then be easily performed 
with the result:
\begin{equation}
\left \langle P_{\rm surv} \right \rangle 
= \frac{1}{1 + \langle t_{\rm age} \rangle/T_{\rm dec}}~.
\label{eq:psurv0}
\end{equation}
Inverting this equation one finds:
\begin{equation}
\frac{\left \langle t_{\rm age} \right \rangle}{ T_{\rm dec}} =
 \frac{1}{\langle P_{\rm surv} \rangle}-1
~.
\label{eq:tage0}
\end{equation}

Before entering the discussion of how to construct a
realistic model for the distribution $f_{\rm age} (t)$ is can be instructive,
for a qualitative understanding of how much the estimate
of $\langle t_{\rm age} \rangle$ is sensitive to the shape
of the distribution, 
to consider a simple generalization of the exponential distribution 
of form:
\begin{equation}
f_{\rm age} (t) = \frac{(1+n)^{1+n}}{\Gamma(1+n)} \;
\frac{t^n}{\langle t_{\rm age}\rangle^{n+1}} \, e^{-t(n+1)/\langle t_{\rm age}\rangle}
\label{eq:form1}
\end{equation}
(an exponential multiplied by a power law of exponent $n$).
This form depends on two parameters, 
the average age  $\langle t_{\rm age}\rangle$
and the adimensional exponent $n$ (that can assume any value in the range
$n > -1$). For $n=0$ the distribution function
reduces to a simple exponential, for $n < 0$ ($n > 0$) the distribution
has an excess of short (long) times. 

Inserting the age distribution (\ref{eq:form1}) 
in equation (\ref{eq:def0}) one obtains:
\begin{equation}
\langle P_{\rm surv}\rangle = \left (\frac{1+n}{1+n 
+ \langle t_{\rm age} \rangle/T_{\rm dec}} \right )^{1+n} ~,
\label{eq:psurv1}
\end{equation}
or inverting
\begin{equation}
\frac{\langle t_{\rm age} \rangle}{T_{\rm dec}} =
\frac{1+n}{\langle P_{\rm surv} \rangle^{1/(1+n)}} - (1+n)
~.
\label{eq:tage1}
\end{equation}
Equations (\ref{eq:psurv1}) and (\ref{eq:tage1}) are the generalizations
of (\ref{eq:psurv0}) and (\ref{eq:tage0}) that are
recovered for $n=0$.
The important point is that the relation between $\langle t_{\rm age} \rangle$,
and the survival probability  $\langle P_{\rm surv} \rangle$
depends strongly on the parameter $n$. For 
large  $\langle t_{\rm age} \rangle/T_{\rm dec}$) 
(and small $\langle P_{\rm surv} \rangle$)
one has the relation
$\langle P_{\rm surv} \rangle \propto \langle t_{\rm age} \rangle^{-(1+n)}$, or
$\langle t_{\rm age} \rangle \propto \langle P_{\rm surv} \rangle^{-1/(1+n)}$,
that is very sensitive to the value of $n$.

More in general, one can observes that the 
relation between $\langle t_{\rm age} \rangle$,
and $\langle P_{\rm surv} \rangle$, is approximately independent
from the shape of the distribution only when
$\langle t_{\rm age} \rangle/T_{\rm dec}$ is small and
$\langle P_{\rm surv} \rangle$ is close to unity.
In fact, 
developing in series the exponential in equation 
(\ref{eq:def0}) one can write $\langle P_{\rm surv} \rangle$ as the series:
\begin{equation}
\langle P_{\rm surv} \rangle = 1 -
\frac{\langle t_{\rm age} \rangle}{T_{\rm dec}} +
\frac{1}{2}
\,\frac{\langle t_{\rm age}^2 \rangle}{T_{\rm dec}^2} + \ldots~.
\label{eq:psurv-series}
\end{equation}
where $\langle t^k_{\rm age} \rangle$ is the $k$--th moment of the distribution.
When the lifetime of the particles is short (and
$\langle P_{\rm surv} \rangle$ is close to unity) one can keep only the 
first term in the series, 
and the relation between $\langle P_{\rm surv} \rangle$
and $\langle t_{\rm age} \rangle$ is unique, but in general 
the dependence of the shape (encoded in the moments $\langle t^k_{\rm age}\rangle$) 
cannot be neglected.
In the concrete situation that has emerged from the observations,
where the quantity $\langle P_{\rm surv} \rangle$ is of order 0.12,
the shape of the residence time distribution
is likely to be very important.

In this work we address the problem outlined above,
constructing a framework to  compute 
the CR residence and escape time distributions
The age distributions can then used to interpret the beryllium 
ratio measurements.
We have  tried to  construct framework that 
is much more realistic than the unphysical leaky--box model,
but that  remains  sufficiently simple 
(it  will contains   only three parameters)
to  yield exact analytic expressions for quantities
of interest.

The work is organized as follows:
in the next section we introduce the simple 1--dimensional diffusion model 
for propagation in the Galaxy, that is the basis of our   calculations,
and compute the shape of the escape time distributions
(that depends on the particle injection point).
In section~\ref{sec:age} we compute the age distributions
(that are a function of the observation point).
In section~\ref{sec:psurv} we compute the average survival 
probability $\langle P_{\rm surv}\rangle$ and study its relation with the 
CR lifetime.
The final section gives a brief summary and some conclusions.

\section{Escape time distribution}
\label{sec:escape}
Diffusion models are in broad use for the description of 
the propagation of cosmic rays in the Milky Way
(see for example \cite{review-CR,Strong:2007nh}. In 
this work we will adopt what can be considered as a ``minimal'' 
diffusion model, where as the CR galactic confinement volume 
one takes all space between the parallel planes $z = \pm z_h$ 
(the subscripts stands for halo).
These boundary planes are considered as absorption surfaces,
and the volume between the planes is considered as filled
by a homogeneous diffusive medium, characterized 
by the isotropic diffusion coefficient $D$.

The assumption of an infinitely large confinement volume is a reasonable 
approximation if the vertical size of the CR halo $z_h$ is much smaller
than the galactic radius. The motivation for introducing this approximation
is that the calculation of the propagation
of cosmic rays is reduced to a 1--dimensional problem 
that has an exact analytic solution.

In general the diffusion coefficient will be
a function of the particle energy (one expects a dependence
of form $D = D_0(|R|) \beta$, with 
$R = p c/(Z e)$ the particle rigidity and 
$\beta$ its velocity), but in this work we will not 
need to specify the functional form of the energy (or rigidity) 
dependence of the diffusion coefficient, because we will 
only discuss the propagation of nuclei,
and assume that the energy of the particles remain
constant after injection without energy loss or reacceleration.
The energy can then be simply considered
as a parameter that labels the propagation of different
particle types. In most of the following discussion
the energy dependence of the different quantities will be left implicit in the
notation.

In our framework the only parameter wih the dimension
of length is the halo half height $z_h$, and by dimensional analysis
one can construct only a single independent quantity with the dimension of time,
the diffusion time
\begin{equation}
T_{\rm diff} = \frac{z_h^2} {2 \, D}~.
\label{eq:tdiff}
\end{equation}
The average escape time and age of the cosmic rays (for a fixed energy)
will of course be proportional to $T_{\rm diff}$, 
times adimensional coefficients that will be calculated below. 

Neglecting energy loss, but allowing for decay and interaction,
the number density $n(\vec{x}, t; E)$ 
of cosmic rays with energy $E$ at the point $\vec{x}$ at the
time $t$ can be obtained from the injection rate
$q(\vec{x}, t; E)$ solving, with appropriate boundary conditions,
the partial differential equation:
\begin{equation}
\frac{\partial n(\vec{x}, t; E)} {\partial t} =
q(\vec{x},t; E) + D(E) \, \nabla^2 n(\vec{x},t; E) 
- \frac{ n(\vec{x}, t; E)} {T_{\rm int} (E)} 
- \frac{ n(\vec{x}, t; E)} {T_{\rm dec} (E)} 
\label{eq:diff1}
\end{equation}
where $T_{\rm int}$ and $T_{\rm dec}$ 
are the interaction and decay time of the particle.
In equation (\ref{eq:diff1}) we have assumed that the interaction rate
of a particle is independent from the position $\vec{x}$.
This implies that the insterstellar gas density is cosidered homogeneous 
in the cosmic ray confinement volume.
In the following we will neglect the effects of interactions
on secondary nuclei. The inclusion, of these effects 
is straightforward if one makes the hypothesis  that 
the gas distribution is homogeneous,  more difficult  in the general case.

The general solution of  the diffusion equation (\ref{eq:diff1}) can be 
expressed in terms of the Green function 
$P(\vec{x}, \vec{x}_i,t)$ that gives the 
probability density that a particle initially at the point $\vec{x}_i$
is at the point $\vec{x}$ after a time $t$:
\begin{equation}
n(\vec{x},t_{\rm obs}) = 
\int_0^\infty dt ~\int d^3 x_i ~q(\vec{x}_i, t_{\rm obs} -t )
~P(\vec{x}, \vec{x}_i, t)
\label{eq:n-integration}
\end{equation}
where $q(\vec{x}, t)$ is the number of cosmic rays
injected per unit time and unit volume at the point $\vec{x}$ at the time $t$.

Neglecting decay and interactions the Green function can be written
explicitely in the form of a series:
\begin{eqnarray}
P_0(\vec{x},\vec{x}_i,t) & = & 
\frac{1}{(4\, \pi \, D \, t)^{3/2}}
~\exp \left[- \frac{(x-x_i)^2 + (y-y_i)^2}{4 \, D \, t} \right ] \times 
\nonumber \\
& ~ & ~ \\ 
& ~& 
 \sum_{n=-\infty}^{+\infty} 
\left \{ \exp \left[ - \frac{(z -z_i - 4 n z_h)^2}{4 \, D\, t} \right ]
- \exp \left[ - \frac{(z +z_i - 2 (2 n+1) z_h)^2}{4 \, D\, t} \right ]
\right \} \nonumber ~.~~~~~~
\label{eq:diffa}
\end{eqnarray}
The function is only defined in the space region $ |z| \le z_h$.
For $|z| \ge z_h$ the particle density and the Green function vanish.

A derivation of equation (\ref{eq:diffa}) is simple.
For propagation in a homogeneous, isotropic diffusive medium
with no boundaries, it is well known that the Green function
is a gaussian of width $\sigma^2 = 2 \, D t$.
In the presence of the two parallel absorption surfaces at $z=\pm z_h$
(see \cite{cox-miller}), the Green function 
becomes the superposition of an infinite number of 
Gaussian functions all of the same width.
One of the gaussians is centered on the real physical source 
at the point with coordinates $\{x_i, y_i, z_i\}$, 
the others correspond to an infinite number of ``mirror''
sinks and sources placed symmetrically outside the diffusion volume.
The sources are located at points $\{x_i,y_i,z_n^+\}$ with:
\begin{equation}
z_n^+ = z_i + 4 \, n \, z_h ~,
\end{equation}
(the source with $n=0$ is the real physical one).
The sinks have 
coordinates $\{x_i,y_i,z_n^-\}$ with:
\begin{equation}
z_n^- = -z_i + 2 \, (2 n+1) \, z_h~.
\end{equation}
In the presence of interaction and/or decay, the Green function becomes: 
\begin{equation}
P(\vec{x},\vec{x}_i,t) =
P_0(\vec{x},\vec{x}_i,t) \; e^{-t/T_{\rm dec}} \;\; e^{-t/T_{\rm int}} ~.
\end{equation}

To compute the escape time distribution 
one can note that the integral of the Green function $P_0$ 
over the entire Galaxy volume ($|z| \le z_h$):
\begin{equation}
{\mathcal N}_0(t, \vec{x}_i) = \int d^3 x ~P_0 (\vec{x}, \vec{x}_i, t) 
\label{eq:nn0}
\end{equation}
gives a result that is unity for $t=0$ and decreases 
monotonically with $t$, vanishing for $t \to \infty$.
This is the consequence of escape, or formally 
absorption at the planes $z=\pm z_h$

The escape time distribution for a particle
that is injected at the point $\vec{x}_i$ can then be calculated as:
\begin{equation}
f_{\rm esc} (t, \vec{x}_i) = - \frac{d{\mathcal N}_0 (t, \vec{x}_i)}{dt}
~.
\label{eq:fesc0}
\end{equation}
The expression is automatically normalized to unity 
(for integration in the range $0 \le t < \infty$).
The space integration and time derivative
in equations (\ref{eq:nn0}) and (\ref{eq:fesc0}) can be easily performed
to obtain an explicit expression for $f_{\rm esc} (t)$.
Our model is effectively unidimensional,
and the escape time distribution depends
only on the $z$ coordinate of the injection point, and has a scaling form:
\begin{equation}
f_{\rm esc} (t, \vec{x}_i) = 
f_{\rm esc} (t, z_i) = 
\frac{1}{T_{\rm diff}} \;
F_{\rm esc} \left (\frac{t}{T_{\rm diff}}, \frac{z_i}{z_h} \right )
\label{eq:fesc1}
\end{equation}
where $T_{\rm diff}$ is the characteristic diffusion time given
in equation (\ref{eq:tdiff}), and the function $F_{\rm esc} (\tau, x)$ is:
\begin{eqnarray}
F_{\rm esc} (\tau, x) &= & \frac{1}{2 \,\sqrt{2 \pi} \,\tau^{3/2}} \;
\sum_{n=-\infty}^{+\infty} ~\left \{ 
e^{- \frac{(x_n^+ +1)^2}{2 \, \tau} } \, (1+ x_n^+) +
e^{- \frac{(x_n^+ -1)^2}{2 \,\tau} } \, (1-x_n^+) 
 \right . \nonumber \\
~& ~& 
\label{eq:ffesc} \\
& ~& ~~~~~~~~~~~~~~\left . 
- e^{ - \frac{(x_n^- +1)^2}{2 \, \tau}} \, (1+ x_n^-) 
- e^{- \frac{(x_n^- -1)^2}{2 \, \tau}} \, (1-x_n^-) 
\right \} \nonumber 
\end{eqnarray}
with $x_n^+ = x + 4 n$
and $x_n^- = -x + 2 (2n+1)$.
For numerical studies the series in equation (\ref{eq:ffesc})
converges very rapidly including the first  few terms with $|n|$ small.

\begin{figure}[t]
\begin{center}
\includegraphics[width=12.7cm]{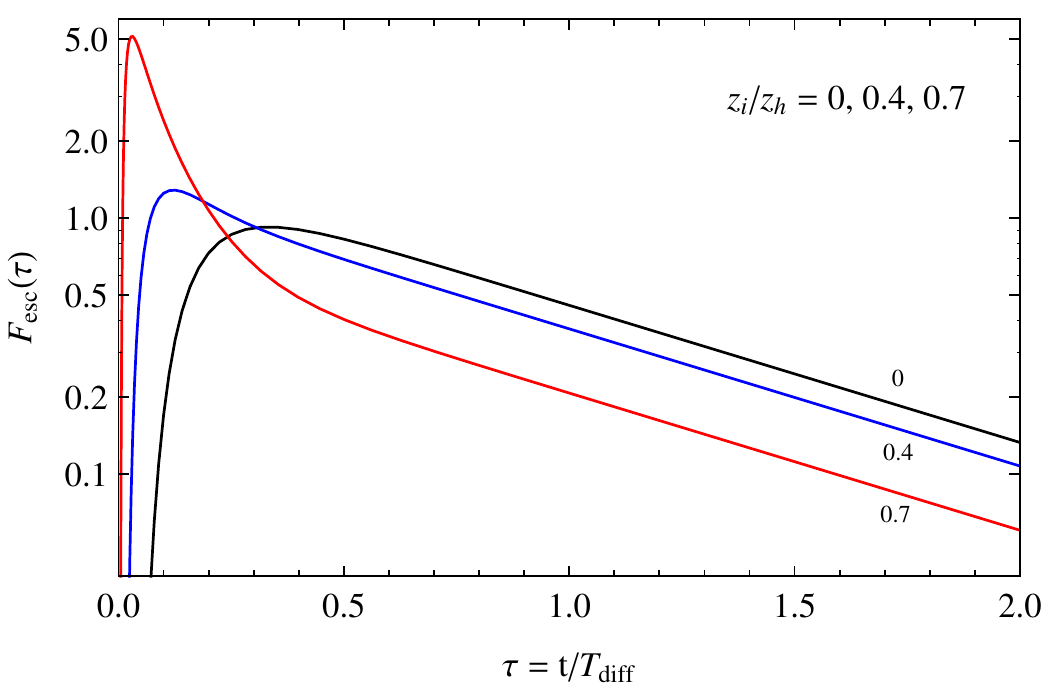}
\end{center}
\caption {\footnotesize
\label{fig:fescape} 
Rescaled escape time distribution
$F_{\rm esc} (\tau,x)$ 
(see equation \ref{eq:fesc1})). The three curves correspond to 
three different injection points ($z_i/z_h = 0$, 0.4, 0.7).
 }
\end{figure}

\begin{figure} [b]
\begin{center}
\includegraphics[width=12.7cm]{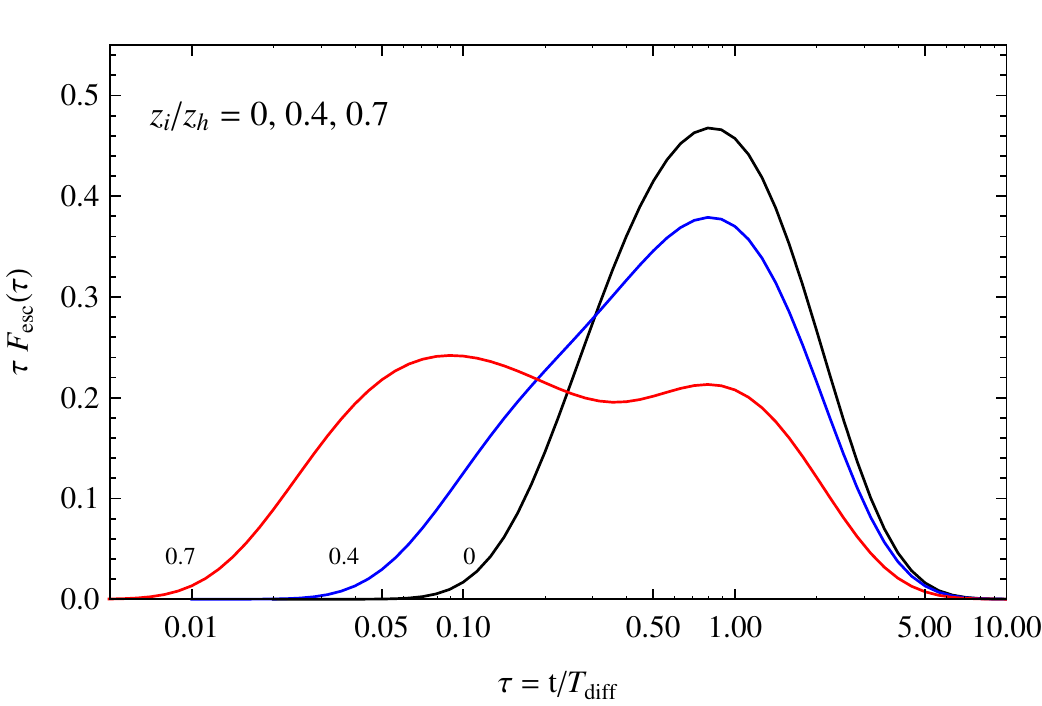}
\end{center}

\caption {\footnotesize
\label{fig:fescape1}
Rescaled escape time distributions $F_{\rm esc} (\tau,x)$ 
(as in fig.~\ref{fig:fescape}). The function is represented in the form 
$\tau \, F_{\rm esc}(\tau)$ versus $\log \tau$. The area under the curves 
is unity.
 }
\end{figure}

Examples of the escape time distribution are shown in 
fig.~\ref{fig:fescape} and fig.~\ref{fig:fescape1}.
Inspecting these figures one can see that for 
large $t$ ($t/T_{\rm diff} \gg 1$) the distribution approaches
asymptotically the exponential form $\propto e^{-t/T^*}$ with a slope
$T^* \simeq 0.810~T_{\rm diff}$ that is independent from the injection 
point.
For small $t$ the distribution vanishes, 
reflecting the fact that the particles
need a finite time to reach one of the boundaries of the diffusion volume
after injection.
The average escape time can be calculated exactly resumming the series,
with the result:
\begin{equation}
\left \langle t_{\rm esc} \left ( \frac{z_i}{z_h} \right) \right \rangle
= \int_0^\infty dt ~t ~f_{\rm esc} (t, z_i) = 
T_{\rm diff} \times \left [1 - \frac{z_i^2}{z_h^2} \right ]
\label{eq:tescav}
\end{equation}
The average escape time is maximum for particles injected 
in the central plane ($z_i=0$), when one has
$ \langle t_{\rm esc} \rangle  = T_{\rm diff}$,
and decreases quadratically with $z_i$, 
when the particle is injected closer to one of the boundaries of the
Galaxy, vanishing for injection at the boundary of the diffusion 
volume ($z_i = \pm z_h$).

\section{Residence time Distribution}
\label{sec:age}

The age distribution for particles at the
point $\vec{x}$ can be obtained from equation (\ref{eq:n-integration}):
\begin{equation}
f_{\rm age} (t, \vec{x}, t_{\rm obs}) 
= \frac{1}{n(\vec{x}, t_{\rm obs})} ~ \int d^3 x_i 
~q(\vec{x}_i, t_{\rm obs} - t) ~P_0(\vec{x}, \vec{x}_i, t)~.
\label{eq:fage0}
\end{equation}
The injection rate $q(\vec{x}_i, t)$ enter in the definition,
so the age distribution is determined also
by the space and time dependence of the injection.

In the following, in the spirit of constructing simple,
exactly solvable models we will consider
an injection that is stationary (independent from $t$) and
has a very simple space dependence.
We will use two models 
for the injection: the ``slab disk'', and the ``exponential disk''.
In both models the injection depends only on the $z$ coordinate,
and is determined by a single parameter (with dimension of length)
that gives the spatial thickness of the emission. 
In the slab disk model the injection 
is homogeneous in the central region of the halo $|z| < z_d$
and vanishes outside:
\begin{equation}
q(z) = \begin{cases}
q 
 & {\rm for}~~ |z| \le z_d \\
 & ~~ \\
 0 & {\rm for}~~ |z| > z_d ~.
\end{cases}
\end{equation}
The parameter $z_d$ can take values in the interval $0 < z_d \le z_h$.

In the exponential disk model the injection has the form
\begin{equation}
q(z) = q_0 ~e^{-|z|/z^*}~,
\end{equation}
where the parameter $z^*$ can take values in the interval
$0 < z^* < \infty$.
The two models are identical when their parameters are at
the extremes of their intervals of definitions, that is for
$z_d = z^* = 0$, when the injection volume is reduced to a plane,
and for $z_d = z_h$ or $z^* \to \infty$, when the injection is
homogeneous in the entire confinement volume.

\begin{figure} [t]
\begin{center}
\includegraphics[width=12.7cm]{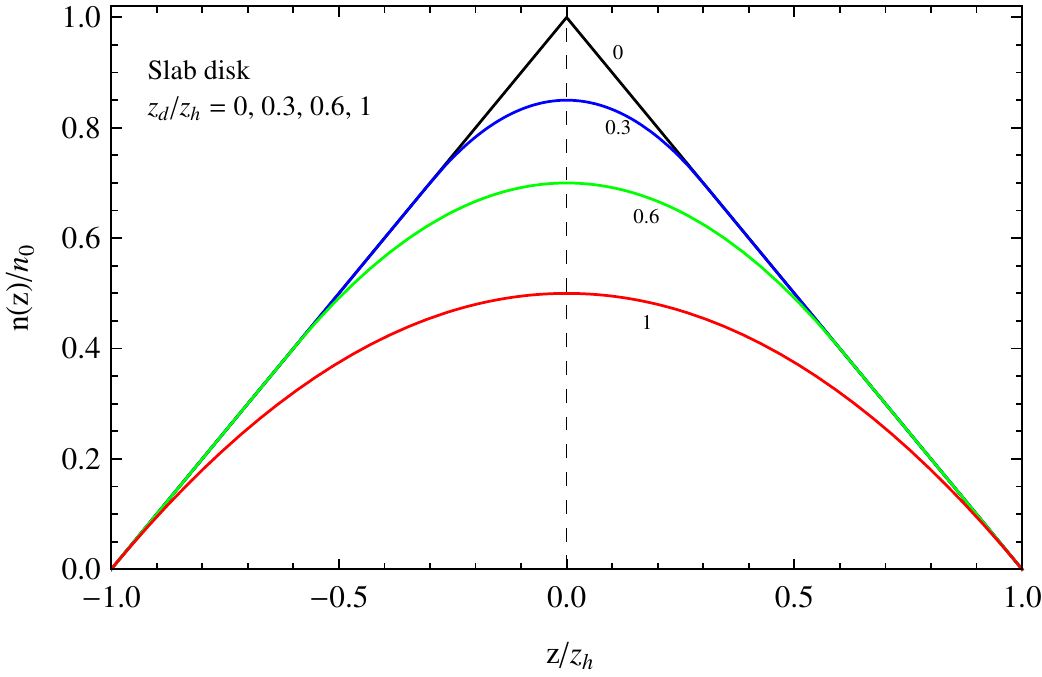}
\end{center}

\caption {\footnotesize
\label{fig:fnslab}
Density of stable cosmic rays in the slab disk injection model 
plotted as a function of the ratio $z/z_h$. The different curves are calculated 
assuming the same total injection rate and different values of the
ratio $z_d/z_h$ ($z_d/z_h = 0$, 0.3, 0.6, 1).
 }
\end{figure}

\begin{figure} [b]
\begin{center}
\includegraphics[width=12.7cm]{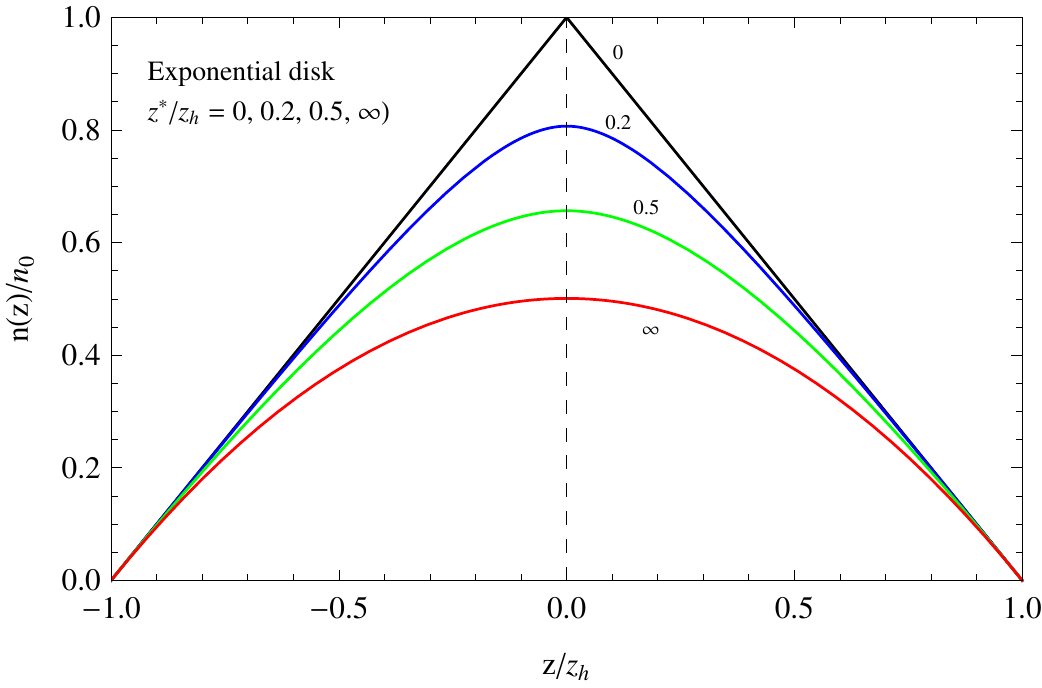}
\end{center}

\caption {\footnotesize
\label{fig:fnexp}
Density of stable cosmic rays in the exponential disk injection model 
model plotted as a function
of the ratio $z/z_h$. The different curves are calculated 
assuming the same total injection rate and different values of the
ratio $z^*/z_h$ ($z^*/z_h = 0$, 0.2, 0.5, $\infty$).
 }
\end{figure}

It is straightforward to compute the density of cosmic rays 
for both models.
In the slab disk injection model one has:
\begin{equation}
n(z; z_d) = \frac{q}{D} \; z_d~z_h
\times 
 \begin{cases}
1 - 
\frac{1}{2} \frac{z_d}{z_h} 
-\frac{1}{2} \frac{z^2}{z_d \;z_h} 
 & {\rm for}~~ |z| \le z_d \\
 & ~~
 \\
1 - \frac{1}{2} \frac{|z|}{z_h} 
 & {\rm for}~~ |z| > z_d ~.
\end{cases}
\label{eq:nslab}
\end{equation}
In the exponential disk injection model:
\begin{equation}
n(z; z^*) = \frac{q_0}{D} \; z^*~z_h
\left [1
-\frac{|z|}{z_h} 
-\frac{z^*}{z_h}
\; \left (
e^{-|z|/z^*} - e^{-z_h/z^*}
 \right )
\right ]
\label{eq:nexp}
\end{equation}
In the models the density
is proportional to the ratio $q/D$ (or $q_0/D$)
and only depends on the $z$ coordinate,
with a shape determined by the adimensional ratios $z_d/z_h$ or $z^*/z_h$,
that goes to zero at the boundaries of the confinement volume 
($|z| \ge z_h$). 
The $z$ dependence of the CR density 
is shown in fig.~\ref{fig:fnslab} and fig.~\ref{fig:fnexp}
for the slab and exponential disk models.

The age distribution can be calculated explicitely using 
equation (\ref{eq:fage0}). The result has the same scaling
property as the escape time distribution
of equation (\ref{eq:fesc1}):
\begin{equation}
f_{\rm age} (t, \vec{x}) = 
f_{\rm age} (t, z) = 
\frac{1}{T_{\rm diff}} \;
F_{\rm age} \left (\frac{t}{T_{\rm diff}}, \frac{z}{z_h} \right )
\label{eq:fage1}
\end{equation}
For the slab disk model one obtains:
\begin{eqnarray}
F_{\rm age} (\tau, x; x_d) &= & 
\frac{1}{2} ~
\sum_{n=-\infty}^{+\infty} ~\left \{ 
 {\rm erf} \left [\frac{x + x_d - 4 n}{\sqrt{2} \, \tau} \right ]
-{\rm erf} \left [\frac{x - x_d - 4 n}{\sqrt{2} \, \tau} \right ]
\right . \nonumber \\
~& ~& 
\label{eq:ftageslab}
\\
& ~& ~~~~~~~~~~~ \left . 
+{\rm erf} \left [\frac{x - x_d - 2(2 n+1)}{\sqrt{2} \, \tau} \right ]
-{\rm erf} \left [\frac{x + x_d - 2(2n+1)}{\sqrt{2} \, \tau} \right ]
\right \} \nonumber 
\end{eqnarray}
where $x=z/z_h$, $x_d = z_d/z_h$ and erf indicates the error function.

For the exponential disk model one has:
\begin{eqnarray}
F_{\rm age} (\tau, x; y) &= & 
\sum_{n=-\infty}^{+\infty} ~\left \{ 
e^{\frac{t + 8 n y 2 x y}{2 y^2}} 
 {\rm erf} \left [
\frac{t + y + 4 n y -x \,y}{\sqrt{2 \, \tau} \, y} 
\right ]
- {\rm erf} \left [
\frac{t + 4 n y -x \,y}{\sqrt{2 \, \tau} \, y} 
\right ]
\right . \nonumber \\
~& ~& 
\label{eq:ftageexp} \\
& ~& ~~~~~~ \left . 
e^{\frac{t + 2(x - 4 n -2) \,y}{2 \,y^2}} 
 {\rm erf} \left [
\frac{t + y (z - 4 n -2)}{\sqrt{2 \, \tau} \, y} 
\right ]
- {\rm erf} \left [
\frac{t + y \,(z - 4 n -1)}{\sqrt{2 \, \tau} \, y} 
\right ]
\right \} \nonumber 
\end{eqnarray}
where again $x=z/z_h$ and $y = z^*/z_h$.

\begin{figure} [t]
\begin{center}
\includegraphics[width=12.7cm]{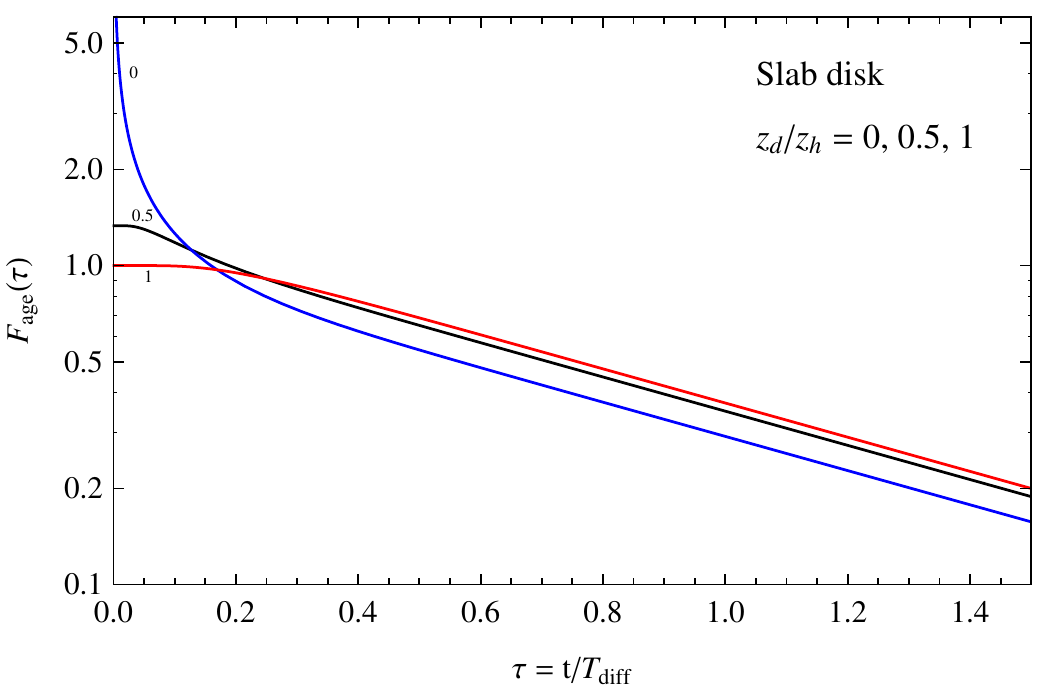}
\end{center}

\caption {\footnotesize
\label{fig:fageslab}
Rescaled age time distribution (see equation (\ref{eq:fage1})).
The function is calculated in the slab disk injection model
for observations at $z=0$. 
The different curves correspond to 
different values of the ratio $z_d/z_h$ ($z_d/z_h = 0$, 0.5, 1).
}
\end{figure}

\begin{figure} [b]
\begin{center}
\includegraphics[width=12.7cm]{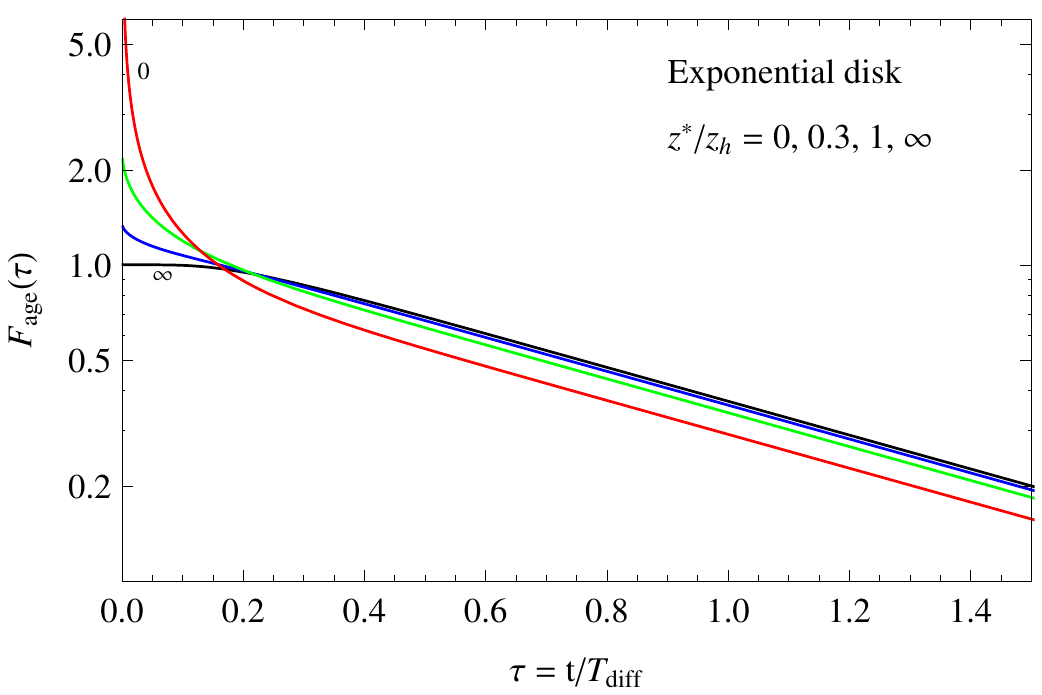}
\end{center}

\caption {\footnotesize
\label{fig:fageexp}
Rescaled age time distribution in the exponential disk model.
The different curves correspond to 
observations at $z =0$ for three 
different values of the ratio 
$z^*/z_h$ ($z^*/z_h = 0$, 0.3, 1, $\infty$).
 }
\end{figure}

Examples of the residence time
distributions are shown 
in fig.~\ref{fig:fageslab} and fig.~\ref{fig:fageexp}.
In these figures the distributions are shown for an observation point
at $z=0$, but for different values of the ratio $z_d/z_h$ or
$z^*/z_h$. The reason to choose a point at $z \simeq 0$ is that 
the observations \cite{Freudenreich:1997bx}
indicate that the solar system lies close (approximately 16~pc) below
the galactic plane.
Inspecting figures~\ref{fig:fageslab} 
and~\ref{fig:fageexp} one can see that the distribution 
for large $t$ takes an asymptotic 
exponential form with slope $T^* \simeq 0.810~T_{\rm diff}$
(the same slope of the escape time distribution at large $t$);
for small $t$ the shape deviates from an exponential
with, in most cases, an excess at short times.
This is the consequence of having a large fraction of the 
sources at short distances from the observation point. 
The excess of small $t$ trajectories
becomes more important when the ratio between the injection and
the confinement volume decreases. In the limit of planar injection
($z_d \to 0$ or $z^* \to 0$ for the two injection models) 
the age distribution at small $t$ diverges as 
$f_{\rm age} (t) \propto 1/\sqrt{t}$.

\begin{figure} [t]
\begin{center}
\includegraphics[width=12.7cm]{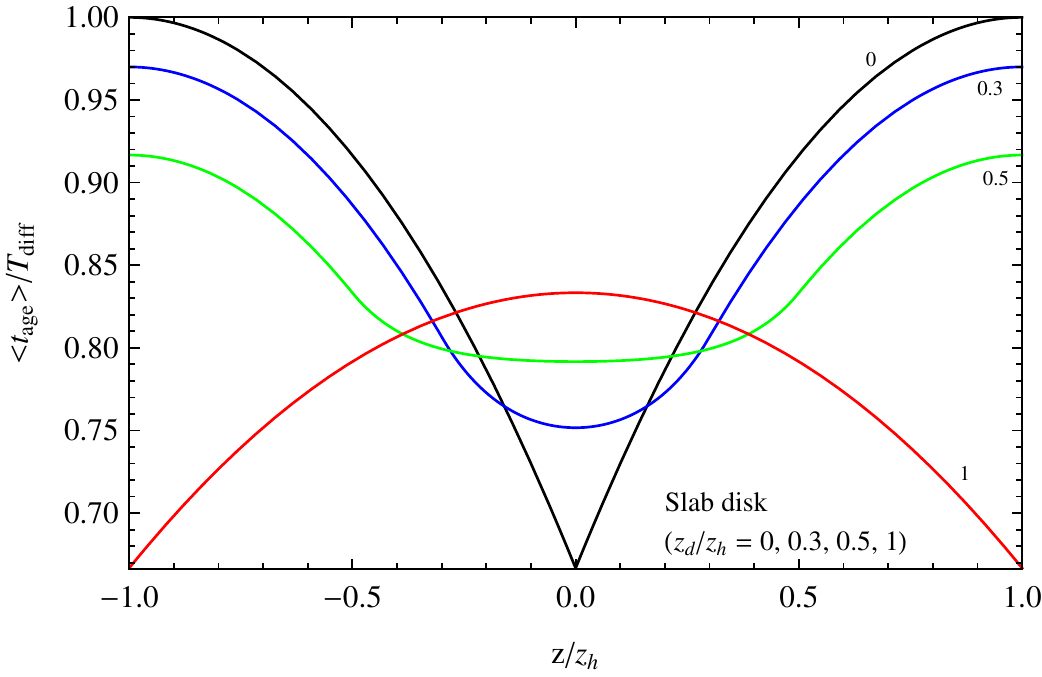}
\end{center}

\caption {\footnotesize
\label{fig:ftageslab}
Average CR residence time (or age) calculated in the slab disk injection model,
plotted as a function of the ratio $z/z_h$. 
The different curves correspond to three values of the ratio
$z_d/z_h$ ($z_d/z_h = 0$, 0.3, 0.5, 1).
 }
\end{figure}

\begin{figure} [b]
\begin{center}
\includegraphics[width=12.7cm]{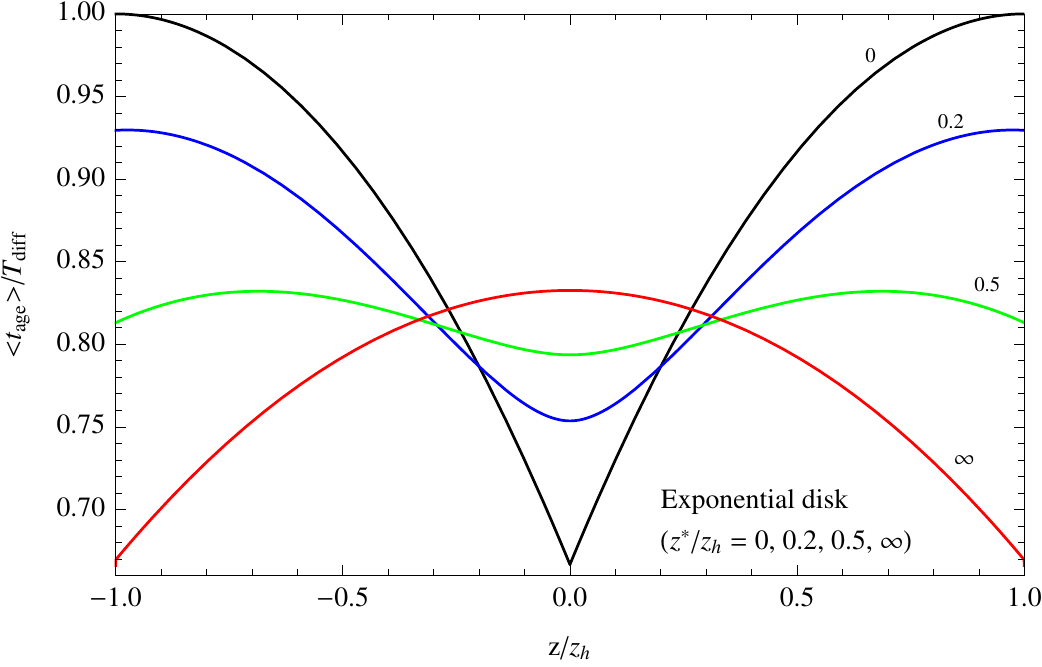}
\end{center}

\caption {\footnotesize
\label{fig:ftageexp}
Average residence time (or age) calculated in the exponential disk 
injection model,
plotted as a function of the ratio $z/z_h$. 
The different curves correspond to three values of the ratio 
$z^*/z_h$ ($z^*/z_h = 0$, 0.2, 0.5, $\infty$).
 }
\end{figure}

The average age can be calculated analytically,
and the result has a simple form. For slab disk injection model one finds:
\begin{equation}
\left .
\frac{\left \langle t_{\rm age} (z,z_d) \right \rangle}{T_{\rm diff}}
\right |_{\rm slab}
= \begin{cases}
\frac{6 x^2\, (2 -x_d) - x_d \, (8 - 4 x_d^2+x_d^3)-x^4}{6[x^2- x_d \, (2 - x_d)]}
 & {\rm for}~~ |z| \le z_d \\
 & ~~ \\
\frac{2}{3} 
+ \frac{2}{3} \, |x| 
- \frac{1}{3} \, x^2 
- \frac{1}{3} \, x_d^2
 & {\rm for}~~ |z| > z_d 
\end{cases}
\label{eq:tageavslab}
\end{equation}
(with $x=z/z_h$ and $x_d = z_d/z_h$).
For the exponential disk injection model:
\begin{equation}
\left .
\frac{\left \langle t_{\rm age} (z,z^*)\right \rangle}{T_{\rm diff}}
\right |_{\rm exp}
= \frac{
e^{\frac{1+x}{y}}(1-x)(2 + 2x - x^2 - 6 y^2)
-3 \, e^{x/y} \, y [1-x^2 - 2 y^2] 
+ 6 \, e^{1/y} y^3}{3 \, 
[e^{\frac{1+x}{y}}(1-x) 
-e^{\frac{1}{y}} \, y
+e^{\frac{x}{y}} \, y]} 
\label{eq:tageavexp}
\end{equation}
(with $x=z/z_h$ and $y = z^*/z_h$).
The average residence time for the two injection models
is shown in fig.~\ref{fig:ftageslab} and fig.~\ref{fig:ftageexp}) plotted
as a function of the $z$ coordinate.

In the limiting cases of planar and volume injections
one obtains simpler expressions. For 
planar injection ($z_d =0$ or $z^* =0$)
one has:
\begin{equation}
\left . \frac{\left \langle t_{\rm age} (z) \right \rangle}{T_{\rm diff}} 
\right |_{\rm plane} =
\frac{2}{3} 
+ \frac{2}{3} \, \frac{|z|}{z_h} 
- \frac{1}{3} \, \frac{z^2}{z_h^2} ~.
\label{eq:tava}
\end{equation}
For homogeneous injection ($z_d = z_h$ or $z^* \to \infty$):
\begin{equation}
\left . \frac{\left \langle t_{\rm age} (z) \right \rangle}{T_{\rm diff}}
\right |_{\rm volume} =
\frac{5}{6} 
- \frac{1}{6} \, 
\frac{z^2}{z_h^2} 
\label{eq:tavb}
\end{equation}
In the first case (injection from a plane) the average CR age is shortest 
when the observation point is at $z=0$, 
($\langle t_{\rm age} \rangle/T_{\rm diff} = 2/3$),
it grows monotonically with $|z|$, with the maximum
at the boundaries of the containement volume at $|z| = z_h$, 
(where $\langle t_{\rm age} \rangle/T_{\rm diff} = 1$).
For homogeneous injection the situation
is reversed, the longest average residence time is at $z=0$ where 
$\langle t_{\rm age} \rangle/ T_{\rm diff} = 5/6$;
the residence time decreases with $|z|$, with minimum 
value at the boundaries
$\langle t_{\rm age} \rangle /T_{\rm diff} = 2/3$.

\section{Average survival probability} 
\label{sec:psurv}
\begin{figure} [t]
\begin{center}
\includegraphics[width=12.7cm]{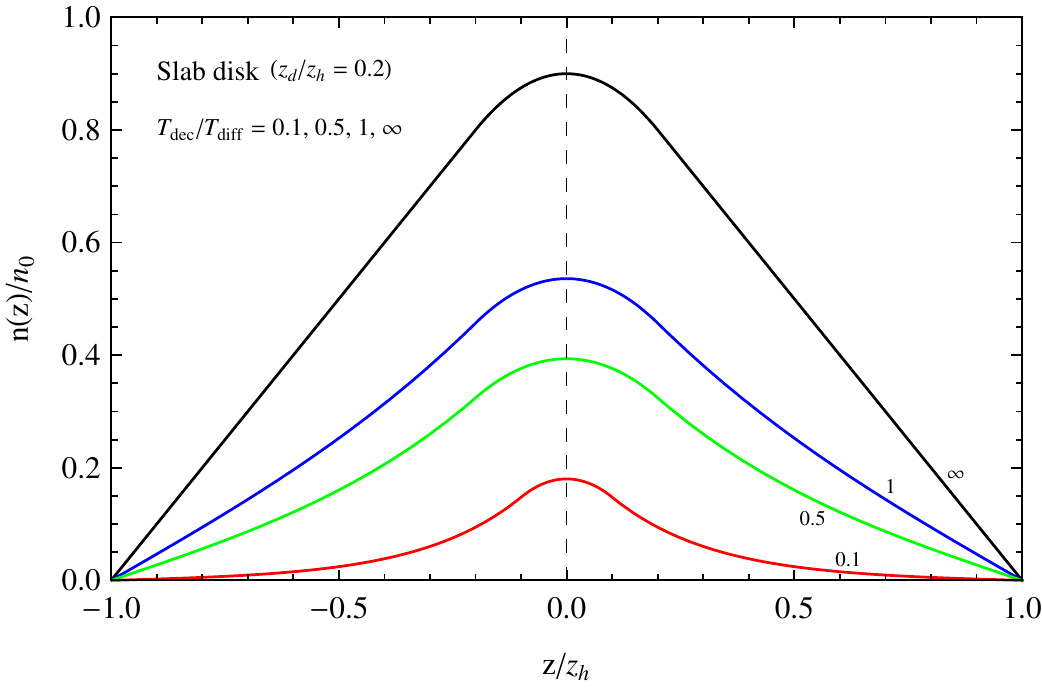}
\end{center}

\caption {\footnotesize
\label{fig:fnslabtime}
Density of unstable cosmic rays in the slab disk injection 
model plotted as a function
of the ratio $z/z_h$. 
All lines are calculated 
for a fixed value of the ratio $z_d/z_h =0.2$. 
The different curves correspond to stable particles
and to unstable particles for a ratio 
$T_{\rm dec}/T_{\rm diff} = 1$, 0.5 and 0.1.
 }
\end{figure}

\begin{figure} [b]
\begin{center}
\includegraphics[width=12.7cm]{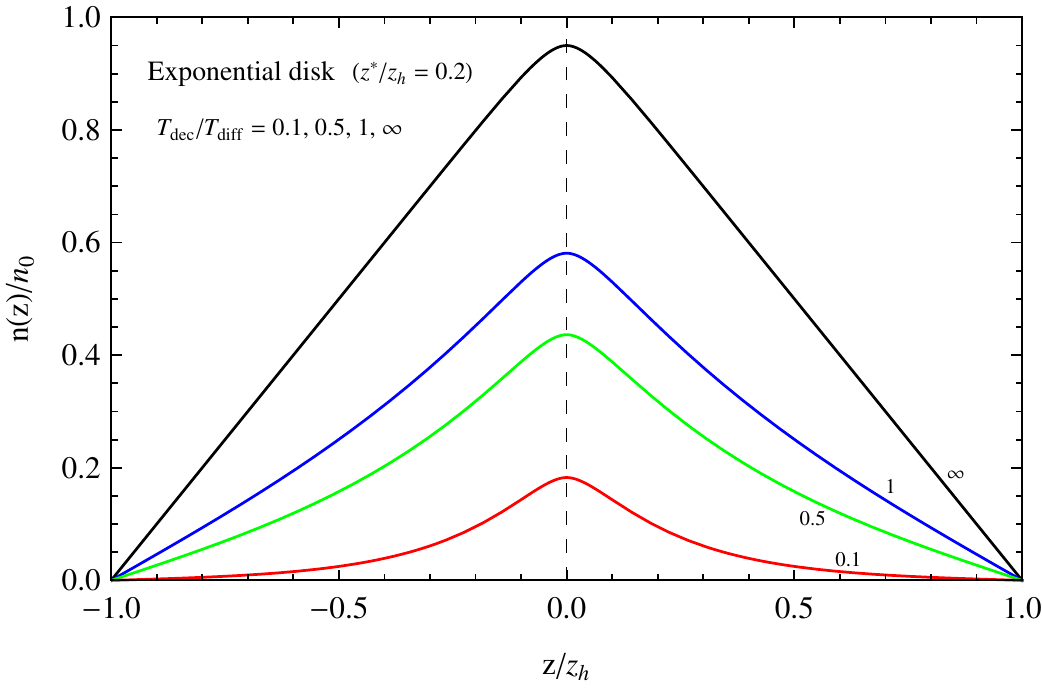}
\end{center}

\caption {\footnotesize
Density of unstable cosmic rays in the exponential disk
injection model plotted as a function
of the ratio $z/z_h$. 
All lines are calculated 
for a fixed value of the ratio $z^*/z_h =0.2$. 
The different curves correspond to stable particles
and to unstable particles for a ratio 
$T_{\rm dec}/T_{\rm diff} = 1$, 0.5 and 0.1.
\label{fig:fnexptime}
 }
\end{figure}

The quantity $\langle P_{\rm surv} \rangle$ can be calculated
from its definition in (\ref{eq:def0}) using the explicit expressions
of the age distributions given in equations 
(\ref{eq:ftageslab}) or (\ref{eq:ftageexp}).
The resulting series can be resummed to obtain a simple analytic expression.
There is however a much simpler method to 
obtain $\langle P_{\rm surv} \rangle$, 
computing the CR density in the entire galactic volume solving equation 
(\ref{eq:diff1}) for an arbitrary value of $T_{\rm dec}$ 
(or of the combination $T^{-1} = T_{\rm dec}^{-1} + T_{\rm int}^{-1}$),
and then taking the ratio with the density calculated 
for a stable particle (that is $T_{\rm dec} \to \infty$):
\begin{equation}
\langle P_{\rm surv} (\vec{x}) \rangle = \frac{n(\vec{x}, T)}
{n(\vec{x},T \to \infty)} ~.
\end{equation}
For the slab disk injection models the particle density is:
\begin{equation}
n(z,T) = q \; T
\times 
 \begin{cases}
\left [
1 - \cosh \left [\frac{z}{\sqrt{D \, T}} \right ]
\cosh \left [\frac{z_h-z_d}{\sqrt{D \, T}} \right ]
\left (\cosh \left [\frac{z_h}{\sqrt{D \, T}} \right ]\right)^{-1}
\right ]
 & {\rm for}~~ |z| \le z_d \\
 & ~~ \\
\sinh \left [\frac{z_d}{\sqrt{D \, T}} \right ]
\sinh \left [\frac{z_h-z}{\sqrt{D \, T}} \right ]
\left (\cosh \left [\frac{z_h}{\sqrt{D \, T}} \right ] \right )^{-1}
 & {\rm for}~~ |z| > z_d ~,
\end{cases}
\label{eq:nslabT}
\end{equation}
For the exponential disk injection model:
\begin{eqnarray}
n(z,T) & = & \frac{q_0 \, T \, z^*} {(z^*)^2 - D \, T} 
\; \left \{ e^{-|z|/z^*} \, z^* 
- e^{-z_h/z^*} \; 
\cosh \left [ \frac{z_h}{\sqrt{D \, T}} \right ]^{-1} \times
 \right . 
\nonumber \\
~ &~& 
\label{eq:nexpT}
\\
& ~& ~~~ 
\; \left . \left ( 
z^* \; \cosh \left [ \frac{z}{\sqrt{D \, T}} \right ] 
+\sqrt{D \, T} \, e^{z_h/z^*} 
\; \sinh \left [ \frac{z_h - |z|}{\sqrt{D \, T}} 
\right ]\right ) \right \} 
\nonumber
\end{eqnarray}
It is straightforward to check that in the limit of
$T \to \infty$ one recovers expressions (\ref{eq:nslab})
and (\ref{eq:nexp}).
Examples of the functions $n(z,T)$ for the two injection
models are shown in fig.~\ref{fig:fnslabtime} and~\ref{fig:fnexptime}.

It is especially interesting to discuss
the average survival probability 
$\langle P_{\rm surv} \rangle$
at $z=0$ because, as already mentioned, the solar system 
is close to the galactic plane.
For $z=0$, in the slab disk injection model one has:
\begin{equation}
\left \langle P_{\rm surv} (\tau, x_d) \right \rangle_{\rm slab} =
\frac{1 - \cosh [\sqrt{2 \tau}(1-x_d)] \;(\cosh [\sqrt{2 \tau}])^{-1}}
{\tau \, x_d (2 - x_d)}~,
\label{eq:psurvslab}
\end{equation}
where $\tau = T_{\rm diff}/T_{\rm dec}$ and $x_d = z_d/z_h$.
In the exponential disk injection model, for $z=0$ one has:
\begin{equation}
\left \langle P_{\rm surv} (\tau, y) \right \rangle_{\rm exp} =
\frac{2 y \sqrt{\tau} \; \cosh [\sqrt{2 \tau}]^{-1} +
e^{1/y} (\sqrt{2} \, \tanh[\sqrt{2 \tau}] - 2 y \sqrt{\tau}}
{2 \sqrt{\tau}\,[e^{1/y} (1- y) -y) (1 - 2 y^2 \tau)]}
\label{eq:psurvexp}
\end{equation}
(with $y = z^*/z_h$).

Equations (\ref{eq:psurvslab}) and (\ref{eq:psurvexp}) can be seen 
as the main result of this work, they give average survival
probability of an unstable particles as a function of the
ratio $T_{\rm diff}/T_{\rm dec}$, for different space distributions
of the injection, and different sizes of the confinement volume.
The quantity $T_{\rm diff}$ is related to the 
average escape time $\langle t_{\rm esc} (z_i) \rangle$ 
by equation (\ref{eq:tescav}) and to the average age 
$\langle t_{\rm age} (z) \rangle$ by equation
(\ref{eq:tageavslab}) or (\ref{eq:tageavexp}).

More in general, the survival probability for observations at
a position $z\ne 0$ can be obtained as the ratio of equations
(\ref{eq:nslabT}) and (\ref{eq:nslab}), or
(\ref{eq:nexpT}) and (\ref{eq:nexp}).

\begin{figure} [t]
\begin{center}
\includegraphics[width=12.7cm]{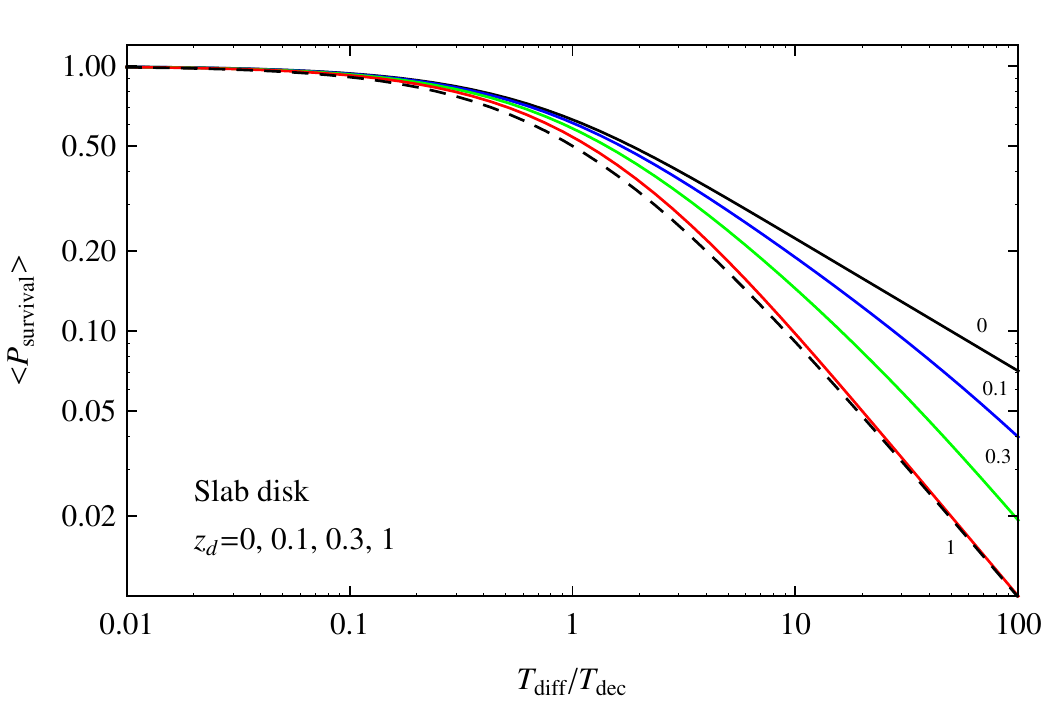}
\end{center}

\caption {\footnotesize
Average survival probability of unstable particles plotted as a
function of the ratio $T_{\rm diff}/T_{\rm dec}$. The different curves
are calculated in the slab disk model for different values
of the parameter $z_d/z_h$ ($z_d/z_h = 0$, 0.1, 0.3, 1).
The dashed line is the leaky--box result plotted as a function
of $\langle t_{\rm esc}\rangle/T_{\rm dec}$.
\label{fig:fpsurvslab}
 }
\end{figure}

\begin{figure} [b]
\begin{center}
\includegraphics[width=12.7cm]{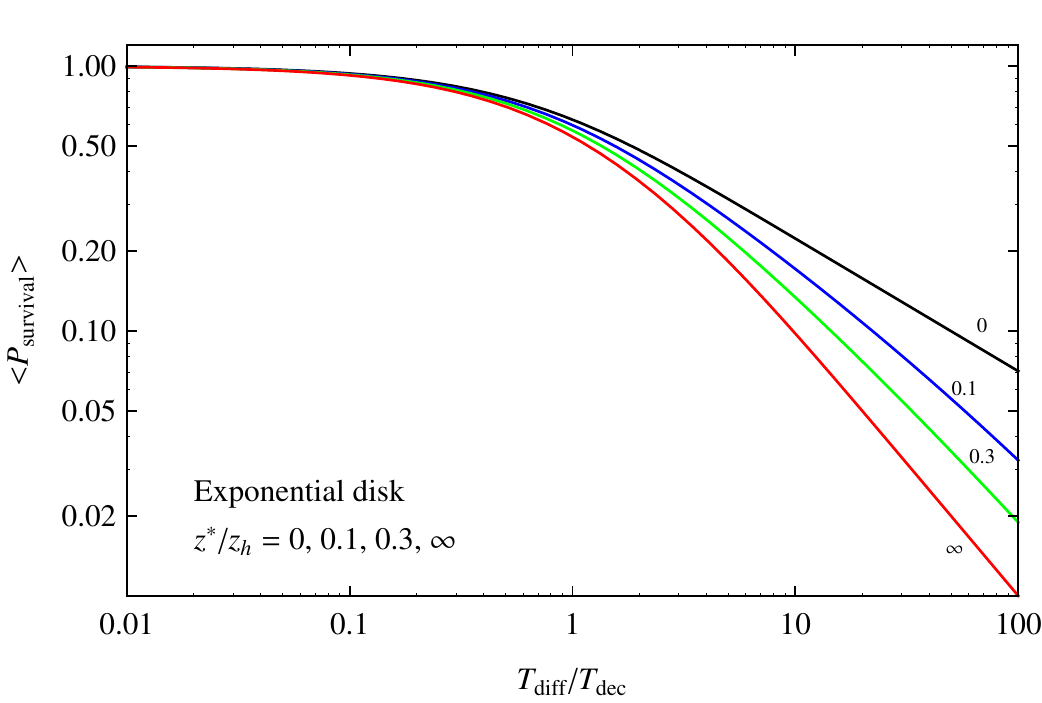}
\end{center}

\caption {\footnotesize
Average survival probability of unstable particles plotted as a
function of the ratio $T_{\rm diff}/T_{\rm dec}$. The different curves
are calculated in the exponential disk model for different values
of the parameter $z^*/z_h$. ($z^*/z_h = 0$, 0.1, 0.3, $\infty$).
\label{fig:fpsurvexp}
 }
\end{figure}

The values of $\langle P_{\rm surv} (\tau) \rangle$ 
for $\tau$ at the two extremes of its range of definition
($\tau=0$ and $\tau \to \infty$) are 
are model independent.
For $\tau \to 0$ (very short residence time) one has
$\langle P_{\rm surv} \rangle = 1$, while 
in the limit $\tau \to \infty$ (very long residence time) 
one has $\langle P_{\rm surv} \rangle = 0$.
In general however, the shape of $\langle P_{\rm surv} (\tau)\rangle$ 
is model dependent.
In the two very simple models discussed here 
the survival probability depends on only one parameter, 
the ratio $z_d/z_h$ for the slab disk injection model,
or the ratio $z^*/z_h$ for exponential disk injection.
These shapes are shown in figures~\ref{fig:fpsurvslab}
and~\ref{fig:fpsurvexp}.

Inspecting the figures
one can see that for large $\tau$
the average survival probability decreases as a power law
($\langle P_{\rm surv} \rangle \propto \tau^{-\alpha}$) with an exponent 
that depends on the ratio $z_d/z_h$ (or $z^*/z_h$).
In the limiting cases of injection from a plane or
from the entire confinement volume 
one finds
(for the observation point at $z=0$):
\begin{equation}
\left \langle P_{\rm surv} (\tau) \right \rangle_{\rm plane}
= \frac{\tanh [\sqrt{2 \tau}]}{\sqrt{2 \tau}}
\label{eq:limplane}
\end{equation}
and
\begin{equation}
\left \langle P_{\rm surv} (\tau) \right \rangle_{\rm volume} =
\frac{1}{\tau} \;
\left ( 1 - \frac{1}{\cosh[\sqrt{2 \tau}]} \right )
\label{eq:limvolume}
\end{equation}
These expressions 
span the entire range of possibilities for the set of models 
discussed in this work, and
for large $\tau$ have the asymptotic behavior:
\begin{equation}
\left \langle P_{\rm surv} (\tau) \right \rangle_{\rm plane}
\to \frac{1}{\sqrt{2 \tau}} 
\end{equation}
and
\begin{equation}
\left \langle P_{\rm surv} (\tau) \right \rangle_{\rm volume}
\to \frac{1}{\tau} ~.
\end{equation}
More in general, the large $\tau$ behavior of the 
survival probability
($\langle P_{\rm surv} \rangle \propto \tau^{-\alpha}$)
has an exponent $\alpha$ in the interval $1/2 \le \alpha \le 1$.

The power law behavior of the average survival probability
for large $\tau$ can be easily understood qualitatively,
as it is related to the shape of the age distribution for small $t$.
In the introduction we have shown
(see equation (\ref{eq:psurv1})) 
 that for an age distribution of form $f_{\rm age} (t) \propto t^n \; e^{-t/T}$ 
(that is $t^n$ for small $t$)
the survival probability has the asymptotic behavior
$\langle P_{\rm surv} \rangle \propto \tau^{-(1+n)}$ (with $\tau = T/T_{\rm dec}$).
In the models discussed here, 
when $t \to 0$ the age distribution $f_{\rm age} (t)$
goes to a constant 
(that corresponds to $\langle P_{\rm surv} \rangle \propto \tau^{-1}$)
for volume injection, 
 and diverges as $t^{-1/2}$ 
(that results in $\langle P_{\rm surv} \rangle \propto \tau^{-1/2}$)
for planar injection.

The different possible functional dependences of
$\langle P_{\rm surv} \rangle$ 
on the ratio $T_{\rm diff}/T_{\rm dec}$ in different models
is reflected in a systematic uncertainty 
in the estimate of the CR residence time,
associated to uncertainties in the size of the
confinement volume.
For $\langle P_{\rm surv} \rangle \lesssim 0.2$, when the use
of the asymptotic expressions is a good approximation,
one has
\begin{equation}
\left . \frac{\overline{\langle t_{\rm esc} \rangle}}{T_{\rm dec}} \right |_{\rm volume}
 \simeq 
\frac{2}{3} \;\frac{1}{\langle P_{\rm surv} \rangle} ~,
\label{eq:est1}
\end{equation}
for volume injection, and
\begin{equation}
\left . \frac{\overline{\langle t_{\rm esc} \rangle}}{T_{\rm dec}} \right |_{\rm plane}
 \simeq 
\frac{1}{2} \;\frac{1}{\langle P_{\rm surv} \rangle^2} ~,
\label{eq:est2}
\end{equation}
for planar injection.
In these equations $\overline{\langle t_{\rm esc} \rangle}$ is the 
particle escape time averaged over all injection points,
and have used equation (\ref{eq:tescav})
to perform the average over all injection points.
For other values of the ratio $z_d/z_h$ or $z^*/z_h$, the 
estimate of the average residence time takes intermediate values between
those given in equations (\ref{eq:est1}) and (\ref{eq:est2}).

\begin{figure} [t]
\begin{center}
\includegraphics[width=12.7cm]{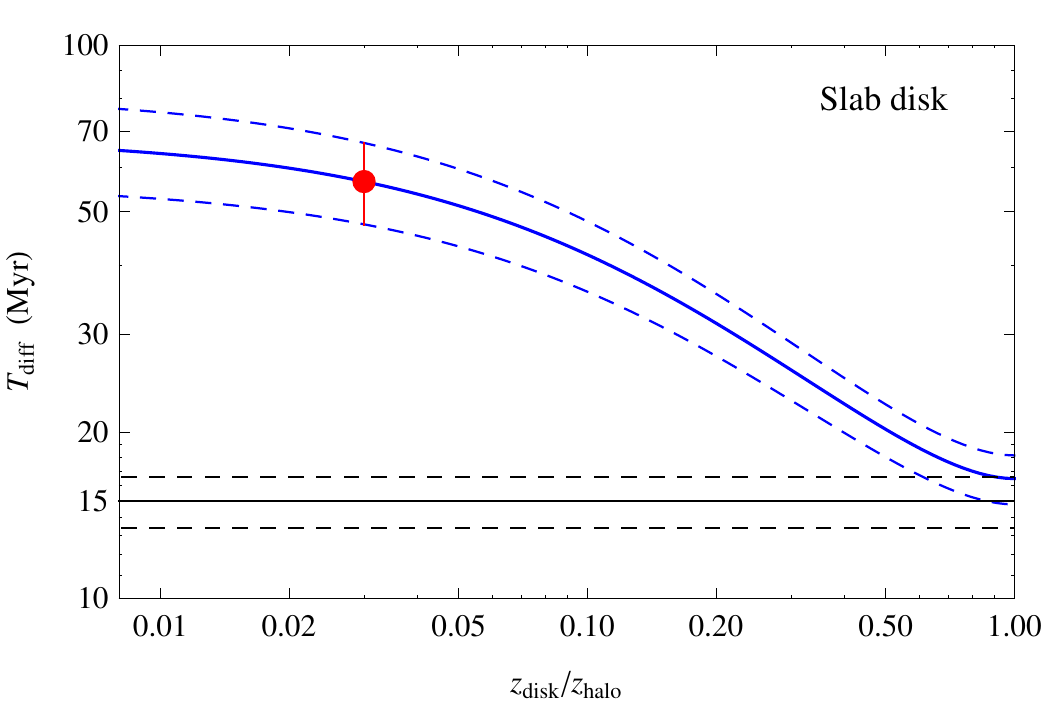}
\end{center}

\caption {\footnotesize
Interval of $T_{\rm diff}$ that corresponds to the 
average survival probability $\langle P_{\rm surv} \rangle$ obtained 
by the CRIS experiment \protect\cite{beryllium3}. 
The interval is calculated for the diffusion model 
discussed in this work with slab disk injection,
and is plotted as a function of the ratio $z_d/z_h$ 
(the ration between the vertical sizes of the source and
confinement volumes). 
The horizontal band is the estimate of $\langle t_{\rm esc} \rangle$ 
estimated by the CRIS collaboration in a leaky--box model
framework.
\label{fig:fintslab}
 }
\end{figure}

\begin{figure} [b]
\begin{center}
\includegraphics[width=12.7cm]{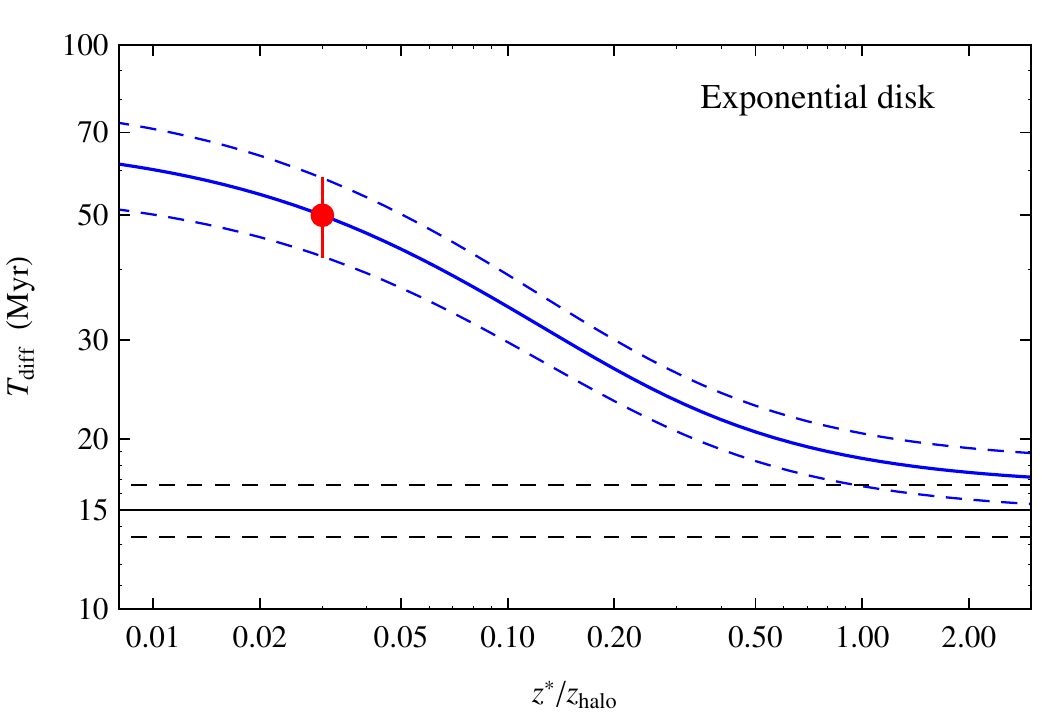}
\end{center}

\caption {\footnotesize 
As in fig.~\ref{fig:fintslab} but the 
$T_{\rm diff}$ interval is calculated for the exponential disk injection model,
and is plotted as a function of the ratio $z^*/z_{\rm halo}$. 
 \label{fig:fintexp} 
 }
\end{figure}

The estimate $\overline{\langle t_{\rm esc} \rangle}$ 
for the volume injection given in 
equation (\ref{eq:est1}) is close to the leaky--box result, 
but the estimate for planar injection
of equation (\ref{eq:est2}) can be much larger.
This is illustrated in fig.~\ref{fig:fintslab} and~\ref{fig:fintexp},
that show the interval of $T_{\rm diff}$ that corresponds to the 
measurement of the beryllium fraction obtained by the CRIS experiment 
 \cite{beryllium3} plotted as a function of 
$z_d/z_h$ or $z^*/z_h$ (for the slab disk and exponential disk injection model).

It is virtually certain that the CR sources are 
in the visible disk of the Galaxy, with a vertical extension
of order 0.1--0.15~Kpc. Therefore in good approximation
the ratio $z_d/z_h$ or $z^*/z_h$ can be interpreted as
the inverse of the vertical extension of the CR halo. 
The observations suggest that the cosmic rays in our Galaxy 
are confined in a volume that is significantly larger than the visible disk,
with a vertical extension of several Kpc. 
This implies that the leaky--box interpretation of the beryllium
isotope ration gives a residence is underestimated
by a factor 2 to 4 for a halo half--thickness between 1 and 5~Kpc.

\section{Conclusions}
\label{sec:conclusions}
The average residence time of cosmic rays in the Milky Way 
is an important quantity that plays a fundamental role
in the estimate of the power of the galactic accelerators.
The best method to determine this residence time is the study of 
the flux of unstable nuclei with a lifetime of appropriate duration.

Measurements of the ratio beryllium--10/beryllium--9 
for nuclei with kinetic energy of order 100~MeV/nucleon
indicate that the flux of the unstable isotope is suppressed
by a factor $\langle P_{\rm surv} \rangle \simeq 0.12$ with respect to
to what is expected in the absence of decay. 
The result clearly shows that the beryllium nuclei 
have a residence time of the same order of the decay time,
but a quantitative estimate requires several assumptions 
about the propagation of cosmic rays in the Galaxy.

The simplest scheme to interpret the beryllium ratio measurements
is the so called leaky--box model
that gives \cite{beryllium3} an average escape time
$15.0 \pm 1.6$~Myr.
The leaky--box model has the merit of a remarkable simplicity,
in fact it can be considered as a ``zero--dimension'' model
where the confinement volume of the cosmic rays is not specified,
and the density and injection of the particles are not allowed
any dependence on the space coordinates.
These unphysical assumptions could (and in fact do) result
in a large bias for the estimate of the CR average residence time,
and it is clearly very desirable to discuss 
the problem in models that are more realistic and have a more
robust physical motivation.

Some rather elaborate numerical codes like GALPROP \cite{Vladimirov:2010aq}
or DRAGON \cite{DiBernardo:2012zu}
have been developed for the study of 
the propagation of cosmic rays in the Galaxy and can be used
to study numerically the confinement of cosmic rays.
In this work we have taken the approach of constructing
a model that is reasonably realistic, but also sufficiently
simple to yield analytic solutions for the quantities of interest,
so that the dependence on the model parameters is  explicit
and transparent, offering a better understanding of the problem.

The framework that we have discussed in this work
can be seen as a ``minimal'' extension of the leaky--box model,
and introduces a minimum number (three) of parameters.
The framework is a one--dimensional diffusion model,
where the Galaxy is an infinite slab of half--thickness $z_h$.
A second parameter determines the space 
distribution of the cosmic ray sources 
(the quantity $z_d$ that gives the half--thickness of the injection volume,
or alternatively the slope $z^*$ for an exponential 
space distribution of the CR injection).
A third parameter (for any given rigidity) is the 
isotropic diffusion coefficient $D$.
The characteristic time for escape from the confinement
volume is then $T_{\rm diff} = z_h^2/(2 D)$.

In this simple framework one can compute exactly the escape time 
and residence time distributions for any possible set of the model parameters.
The age and escape time 
distributions are distinct from each other and are not
unique because they depend 
on the coordinates of the particle injection point (for the escape time) 
or observation point (for the age time).
The average values
 $\langle t_{\rm esc} (\vec{x}_i) \rangle$ 
and
 $\langle t_{\rm age} (\vec{x}) \rangle$ are proportional to $T_{\rm diff}$ 
times a space--dependent, adimensional coefficient that depends only on
the ratio $z_d/z_h$ or $z^*/z_h$.
The distinction between escape time and age does not exist
in the leaky--box model, but is important in
a more general discussion of the cosmic ray residence time.

The most interesting result we have obtained is a simple,
closed form expression for the average survival probability
of an unstable particle $\langle P_{\rm surv}\rangle$ 
that depends on the $z$ coordinate of the observation point,
and is a function of the ratio $T_{\rm diff}/T_{\rm dec}$
(with $T_{\rm dec}$ the decay time of the unstable particle).
The crucial point is that 
$\langle P_{\rm surv}\rangle$ also depends on 
the ratio $z_d/z_h$ (or $z^*/z_h$) between the
injection and confinement volumes.

The bottom line is that given a measurement of 
 $\langle P_{\rm surv}\rangle$ that gives
the flux suppression for an unstable nucleis,
the estimate of the characteristic time $T_{\rm diff}$ 
(proportional to the average escape time and age of the
cosmic rays) depends on the ratio between the
injection and confinement volumes for the cosmic rays or,
assuming that the injection volume is known, only on
the size of the cosmic ray halo.

If the injection and confinement volumes are
approximately equal, the relation between $T_{\rm diff}$ and
$\langle P_{\rm surv} \rangle$ is very close to 
what is estimated in the leaky--box model,
but when the confinement volume (or $z_h$) grows,
the estimate of $T_{\rm diff}$ (for a fixed value of
$\langle P_{\rm surv} \rangle$) increases monotonically.
For a vertical size of the galactic CR halo of order $\simeq 5$~Kpc,
the beryllium isotope ratio measurements imply a lifetime of order
50--60 million years, approximately a factor four longer than 
the leaky--box result.
This corresponds to an equal reduction of the power of the galactic CR sources.

\clearpage


\clearpage


\clearpage

\clearpage

\clearpage

\clearpage

\begin{thebibliography}{100}

\bibitem{Tilley:2004zz} 
 D.~R.~Tilley, J.~H.~Kelley, J.~L.~Godwin, D.~J.~Millener, J.~E.~Purcell, C.~G.~Sheu and H.~R.~Weller,
 Nucl.\ Phys.\ A {\bf 745}, 155 (2004).


\bibitem{beryllium1} 
M.~Garcia-Munoz, G.M.~Mason \& J.A.~Simpson 
``The age of the galactic cosmic rays derived from the abundance of Be-10''
Astrophys. \ J. {\bf 217}, 859 (1977).

\bibitem{beryllium2} 
S.P.~Ahlen {\it et al.}
``Measurement of the Isotopic Composition of Cosmic-Ray Helium, Lithium, Beryllium, and Boron up to 1700 MEV per Atomic Mass Unit''
Astrophys. \ J. \ {\bf 534}, 757 (2000).

\bibitem{beryllium3} 
N.E.~Yanasak {\it et al.}
``Measurement of the Secondary Radionuclides 10Be, 26Al, 36Cl, 54Mn, and 14C and Implications for the Galactic Cosmic-Ray Age''
Astrophys.\ J.\ {\bf 563}, 768 (2001).


\bibitem{review-CR} V.~Ginzburg, V.~Dogiel, V.~Berezinsky, S.~Bulanov, 
and V.~Ptuskin ``Astrophysics of Cosmic Rays'',
North-Holland, Amsterdam, (1990).

\bibitem{Strong:2007nh} 
 A.~W.~Strong, I.~V.~Moskalenko and V.~S.~Ptuskin,
 Ann.\ Rev.\ Nucl.\ Part.\ Sci.\ {\bf 57}, 285 (2007)
 [astro-ph/0701517].

\bibitem{cox-miller}
D.R. Cox and H.D. Miller,
``The Theory of Stochastic Processes'',
Chapman and Hall (1965).

\bibitem{Freudenreich:1997bx} 
 H.~T.~Freudenreich,
 Astrophys.\ J.\ {\bf 492}, 495 (1998)
 [astro-ph/9707340].

\bibitem{Vladimirov:2010aq} 
  A.~E.~Vladimirov {\it et al.},
  Comput.\ Phys.\ Commun.\  {\bf 182}, 1156 (2011)
  [arXiv:1008.3642 [astro-ph.HE]].


\bibitem{DiBernardo:2012zu} 
  G.~Di Bernardo, C.~Evoli, D.~Gaggero, D.~Grasso and L.~Maccione,
  JCAP {\bf 1303}, 036 (2013)
  [arXiv:1210.4546 [astro-ph.HE]].

\end{thebibliography}
\end{document}